# Data-Driven Electron Microscopy: Electron Diffraction Imaging of Materials Structural Properties

Jian-Min Zuo[a,b], Renliang Yuan[a,b], Yu-Tsun Shao[a,b], Haw-Wen Hsiao[a,b], Saran Pidaparthy[a,b], Yang Hu[a,b], Qun Yang[c], Jiong Zhang[d]

[a] Department of Materials Science and Engineering, University of Illinois at Urbana-Champaign, Urbana, IL 61801, USA

[b] Materials Research Laboratory, University of Illinois at Urbana-Champaign, Urbana, IL 61801, USA

[c] School of Physical Science and Technology, ShanghaiTech University, Shanghai, 201210, China

[d] Intel Corporation, Corporate Quality Network, Hillsboro, OR 97124, USA

**Abstract**

Transmission electron diffraction is a powerful and versatile structural probe for the characterization of a broad range of materials, from nanocrystalline thin films to single crystals. With recent developments in fast electron detectors and efficient computer algorithms, it now becomes possible to collect unprecedently large datasets of diffraction patterns (DPs) and process DPs to extract crystallographic information to form images or tomograms based on crystal structural properties, giving rise to data-driven electron microscopy. Critical to this kind of imaging is the type of crystallographic information being collected, which can be achieved with a judicious choice of electron diffraction techniques, and the efficiency and accuracy of DP processing, which requires the development of new algorithms. Here, we review recent progress made in data collection, new algorithms, and automated electron DP analysis. These progresses will be highlighted using application examples in materials research. Future opportunities based on smart sampling and machine learning are also discussed.

# 1. Introduction

The background of this review is atomic structure determination using transmission electron microscopy (TEM). In current practice, such analysis is primarily performed using atomic resolution imaging. At atomic scale, aberration corrected (scanning) TEM ((S)TEM) now routinely achieves sub-Å resolution for imaging single atoms [1, 2], defects in a diversity of hard materials [3-5], from ultra-thin 2D atomic layers [6-8] to nanoparticles and crystals [9-11] and more recently the structure of porous materials [12]. The versatility of STEM, especially, is attractive for direct imaging of local atomic structures, including three-dimensional (3D) structure [3, 13-15]. However, the atomic structure information obtainable by atomic resolution imaging is still limited to small volumes [15] and mostly to crystals [9].

An emergent solution to extend atomic structure study to a broad range of materials and length scales is diffraction imaging, or diffraction mapping, that uses recorded DPs for imaging [16-27]. STEM, in principle, is a diffraction imaging technique, however, its applications are limited by detectors. By collecting large DP datasets in a TEM for every probe position and forming diffraction images, the structure of materials can be interrogated at multiple levels. For example, at Å scale, ptychography achieves the spatial resolution beyond the diffraction information limit [18, 21]. At nm to micron scales, orientation, phase and strain mapping provides critical information about microstructure, including lattice rotation and deformation in the presence of defects and interfaces [22, 23, 25, 26, 28]. Electron diffuse scattering recorded in DPs also provides information about local disorder, from short-range order (SRO) to the presence of dislocations and stacking faults [11, 29, 30]. Ultimately, grain boundary and orientation imaging at mesoscopic scale can be combined with atomic and nanoscopic structural information for a complete determination of materials' atomic structure.

The versatility of electron diffraction imaging comes from the location specificity of collected DPs, the flexibility that comes with computer post-processing and data mining, and the information provided by the fundamentals of diffraction physics. In detail, a large dataset is collected in the form of $I(S,x,y)$ with $I$ for diffraction intensity, $x$ and $y$ for the detector coordinates, and $S$ for the sampling points. Using computer processing, the diffraction signals are transformed into crystallographic information for imaging (Figure 1). This data-driven approach to electron microscopy has unique advantages over the traditional microscopy of taking pictures in a (S)TEM. First, Bragg diffraction analysis can be highly quantitative by taking account of dynamical diffraction effects [31, 32]. Second, electron diffuse scattering recorded in a nanodiffraction pattern can be used to study local disorder. Third, an assortment of diffraction techniques can be incorporated in electron diffraction imaging for data collection. Techniques, such as scanning electron nanodiffraction (SEND) [33-35], four-dimensional STEM (4D-STEM) [19, 36], scanning precession electron diffraction (SPED) [37] and scanning convergent beam electron diffraction (SCBED) [38], can be used to collect different types of DPs, from ronchigrams to rocking curves to spot DPs. For data mining, a variety of methods can be brought to bear for the extraction of crystallographic information and for the integration of current knowledge with advancements in data science and machine learning (ML).

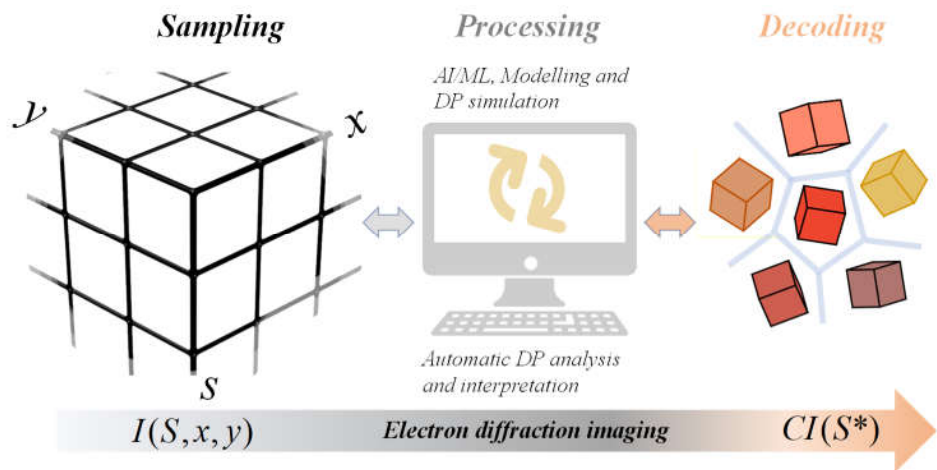

*Figure 1 The concept of electron diffraction imaging, from data collection to processing to image decoding (reconstruction). CI stands for crystallographic information.*

Here we review major progresses that have been made towards electron diffraction imaging with focus on crystallographic information imaging (CII). This review will begin with an introduction to the basic concept of diffraction imaging and its methods, followed by discussions on diffraction imaging modes and applications. For the last part, we highlight the major progress in applying electron CII methods to the characterization of modern materials. We finish with a brief look into future opportunities.

## 2. Experimental Methods

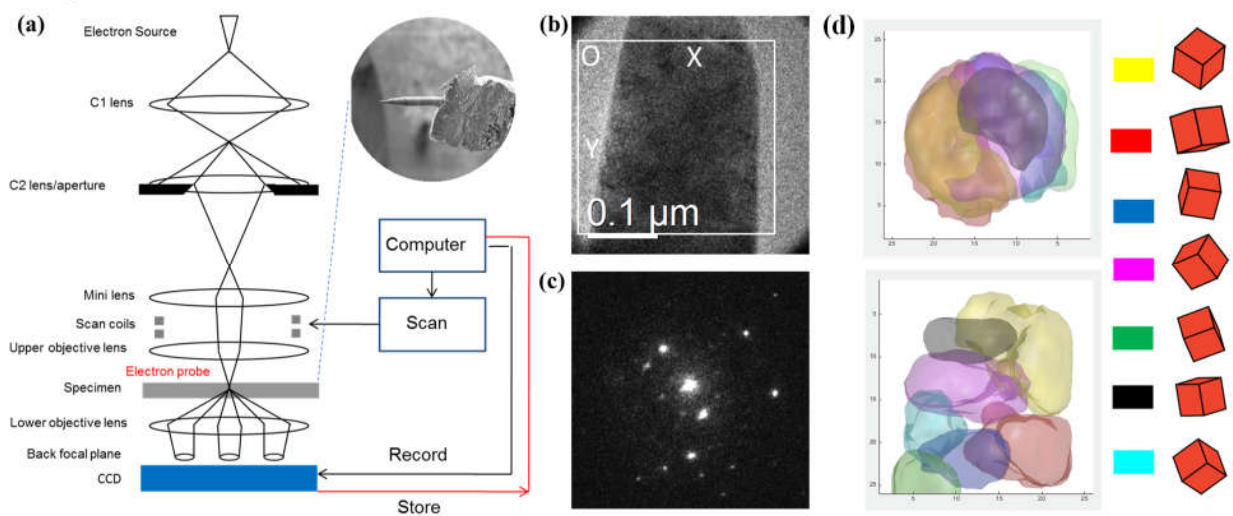

*Figure 2 An example of electron CII. a) Schematic diagram of electron nanoprobe formation and the setup for scanning electron nanodiffraction (SEND). The computer system is used to position the probe, acquire, and store electron diffraction patterns. b) Image of TiN nanotip with the white rectangle indicates the area covered by beam scanning. (c) A sample experimental DP acquired during the beam scanning. d) Reconstructed tomograms of nanograins and their orientation in the TiN nanotip. (From Ref. [39])*

The principle of electron diffraction imaging is demonstrated in Figure 2 using the determination of TiN nanostructure reported by Meng and Zuo [39] as an example. A large dataset of DPs is collected at each rotation angle in a region of interest (ROI) covering the needle-shaped sample using SEND. The DPs were then processed to determine the crystal grains and grain orientations. The processing involves spot DP clustering for grain identification, DP indexing, 3D reconstruction and orientation determination of identified grains. The crystallographic information dataset, $CI(S^*)$, where $CI$ in this case is the crystal orientation and $S^*$ represents the transformed sample coordinates, is then used to construct the orientational tomograms, where the orientation color index is rendered on the 3D Cartesian coordinates (Figure 2d).

## 2.1 Electron Nanodiffraction

The first step in electron diffraction imaging is collecting DPs over a region of interest (ROI). Traditionally, transmission electron diffraction was performed using a broad beam illumination for selected area electron diffraction. A focused beam was used for CBED, which is also known as electron microdiffraction [40]. In electron nanodiffraction, a nm-sized electron beam is used for diffraction. The history of electron nanodiffraction started when the nm-sized electron beams became first available with the development of dedicated STEMs in 1980s [41-45]. However, these instruments had an illumination system designed largely for STEM imaging. Performing nanodiffraction using conventional TEMs at that time was difficult. Major technological innovations were introduced in 1990s that changed this situation. These innovations included the condenser mini-lens and Köehler illumination (for a review, see [38]). Significant improvement was also made in the smallest electron probe with the development of probe aberration correctors, which is now down to sub-Å with a large beam convergence angle [46]. Together, these technologies, now embodied in a modern STEM, enable the formation of a range of electron beams with probe size from sub-Å to tens of nm for electron nanodiffraction.

The small electron beam formed in a STEM with a field emission gun (FEG) is highly coherent and the beam convergence angle plays a critical role in how and where electron interferes and how the interference impacts on DPs. Figure 3a and b show the amplitude and phase diagrams of a simulated coherent probe. The amplitude profile shows the characteristic primary peak at the center and the smaller secondary peaks on both sides, while the phase increases by $\pi$ for each secondary peak (Figure 3c). The complex probe wave function can be divided into concentric zones, each zone has either the phase of 0 or $\pi$. Each zone contributes a part to the scattered wave. The extent of crystal potential seen by the electron probe is approximately determined by the size of first zone ($r_o$), whose radius in absence of aberration is $r_o = 0.61\lambda/\theta_c$. Where $\theta_c$ is the beam semi-convergence angle. As $r_o$ increases and convergence angle decreases, DPs change from ronchigram to CBED patterns, and then to nanodiffraction patterns. These three types of DPs provide crystallographic information for atomic scale, sub-nm and nm scales, which serve as the foundation for electron diffraction imaging.

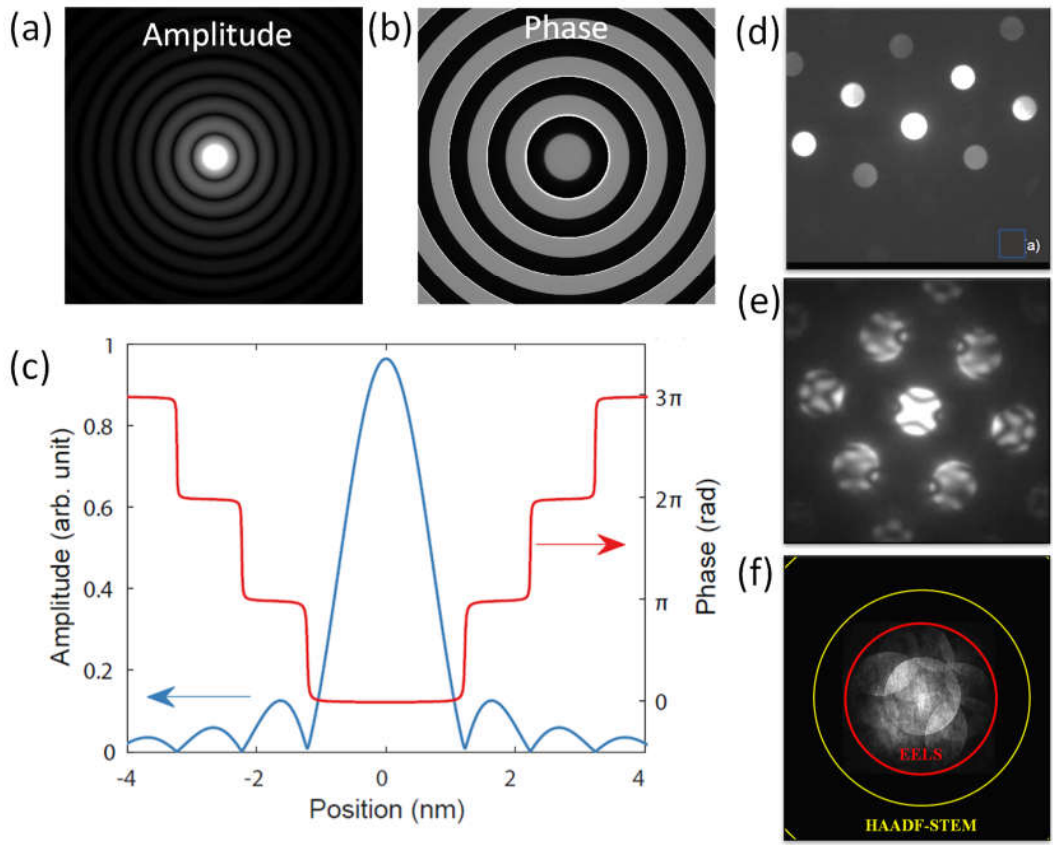

*Figure 3 Electron probe complex wavefunction and diffraction zones. (a) and (b) the amplitude and phase of a simulated nm-sized coherent electron probe. (c) Profiles through the center of the probe. (d), (e) and (f) types of electron diffraction patterns from small to medium to large convergence angles for Si [110].*

## 2.2 SEND, Fast Detectors and 4D-STEM

The collection of DPs is done by combining beam scanning with digital recording using a pixelated electron detector. The early adoptions of SEND often used a charge-coupled device (CCD) camera [22, 33, 47-51], and in some case mounted outside a TEM with a small camera format [35, 52]. Slow-scan CCD cameras, which played a critical role in the digital revolution of TEM, provide a combination of detector quantum efficiency (DQE), dynamic range, and resolution for electron imaging and diffraction [53]. However, their major drawback is the slow readout. To address this, an assortment of faster pixelated electron detectors has been developed, creating new opportunities for electron diffraction. This rapidly evolving field was recently reviewed [54, 55]. Two major solutions have emerged. One is based on Complementary Metal Oxide Semiconductor (CMOS) sensors. The Active Pixel Sensor (APS) designed in CMOS converts signals to a voltage at the pixel site, which is read out using on-chip digital-to-analog converters (DACs). The other solution is to use hybrid direct electron detectors, such as Electron Microscope Pixel Array Detector (EMPAD) and Medipix, which provide significant improvements in the readout speed (1000 frames per second or faster), as well as in DQE and the detector dynamic range [54, 56-58]. New CMOS APS with very high readout speed is also under development [59]. Figure 4 shows two example DPs recorded using EMPAD with a sensor format of 128 by 128 pixels [56], demonstrating the quality of DPs for a single frame.

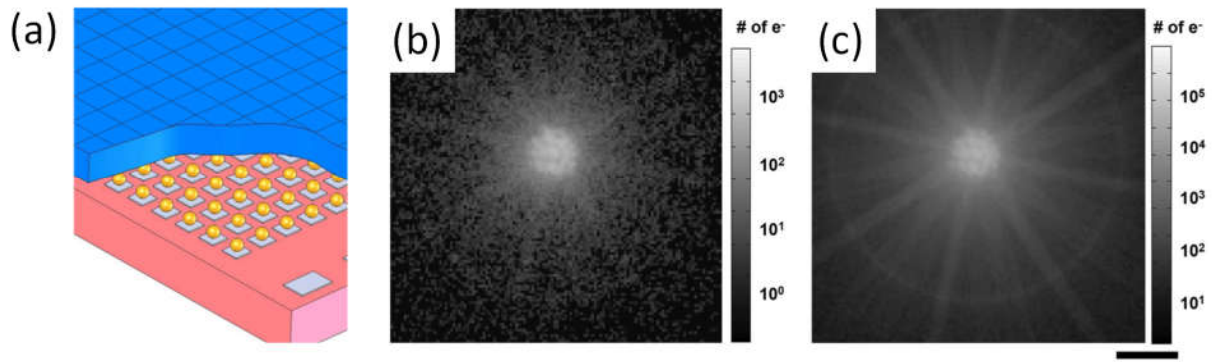

*Figure 4 Hybrid electron direct detection detector. (a) Schematic illustration showing the pixelated Si diode absorber layer (blue) and a integrated circuit device for read out. (b) and (c) CBED pattern of $BiFeO_3$ recorded in 1 ms and 100 ms with 10 pA of beam current at a single scan position, respectively. The scale bar is in unit of the number of primary electrons detected. Black bar on the lowermost far right represents 20 mrad for the diffraction patterns above. (From Tate et al.[56])*

The development of fast direct detection cameras was also motivated by the growing interest in replacing conventional STEM detectors with fast pixelated detectors [36, 56, 58, 60]. By recording electron DP directly, rather than integrating diffraction intensities as done using a conventional STEM detector, a great amount of flexibility is achieved in STEM imaging. This technique became known as 4D-STEM [19], , where the 4D phase space consists of real space probe positions (x,y) and diffraction space ($k_x$, $k_y$).

## 2.3 Precession, Energy Filtering and EBSD

Other major developments that benefit electron diffraction imaging are electron energy filtering and precession electron diffraction (PED). Electron imaging energy-filters were developed to enable the inelastic background from plasmon, or higher electron energy losses, to be removed from recorded DPs [61]. This greatly facilitates the analysis of electron diffraction intensities [30]. Precession electron diffraction (PED) was developed by Vincent and Midgley [62-65]. This technique helps electron crystallographic analysis by reducing the effect of multiple scattering in the measured diffraction intensities and by increasing the number of diffraction spots for DP indexing.

Electron backscattering diffraction (EBSD) in a scanning electron microscope (SEM) is a widely used technique for microstructure determination. Through automatic indexing of Kikuchi patterns captured by EBSD, information about crystal orientation and phase is obtained and mapped, which allows a determination of grain sizes, shapes and grain/phase boundaries [66]. Compared with the analysis of Kikuchi lines, the analysis of transmission electron DPs is more challenging as the recorded DPs contain a variety of features dependent on the sample thickness and the electron diffraction technique employed, as the examples in Figure 3d-f show. TEM does offer several major advantages over a SEM. Chief among them are the electron penetration power. The high spatial resolution and the range of information that can be retrieved using a TEM are also very attractive for studying local compositional order, crystal orientation, lattice strain and distortion and bonding. While X-ray and synchrotron diffraction provide even more penetration power, the advantage of electron diffraction is the large scattering cross-sections of high energy electrons and the small probe size.

## 3. Diffraction Imaging Modes and Applications

Various imaging modes have been developed and applied in electron diffraction imaging. The processing and analysis steps involved in these imaging modes range from being relatively simple, such as the STEM related imaging modes of annual dark-field (ADF), bright-field (BF) and different phase contrast (DPC) [67], to complex, such as ptychography. The STEM related imaging modes were recently reviewed by Ophus [19], while electron ptychography has been reviewed by Rodenburg [21]. Here, we focus on imaging modes that have been developed for CII and their applications to crystalline materials. For the study of noncrystalline materials using electron diffraction, see the review by Voyles and Hwang [68].

### 3.1 Weak-Beam Diffraction Imaging (WBDI)

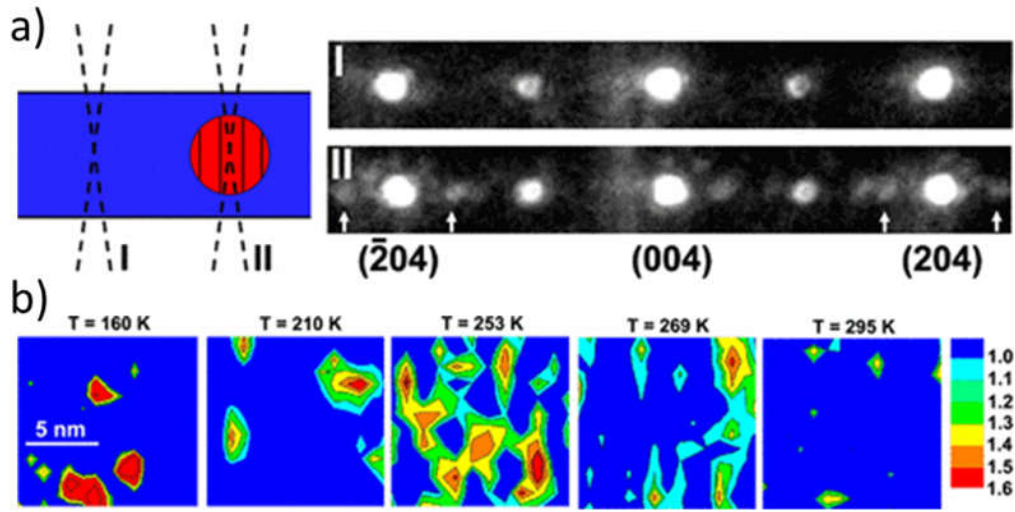

*Figure 5 Imaging charge ordering nanoclusters using weak superlattice reflections. a) The principle of diffraction imaging of an embedded CO cluster and DPs along the [010] zone of $La_{0.55}Ca_{0.45}MnO_3$ from regions I and II, respectively. b) Intensity maps of the marked superlattice reflections (SLR) in II as function of temperature. A color scale of relative intensity is shown on the right. The flat blue area has the intensity of SLRs below the noise level. Each map is from an area of $12\times12$ $nm^2$ with the scanning step of 1 nm, although each image is not necessarily in precise registration with any other (from Tao et al. [33]).*

The use of weak diffraction signals was one of the major motivations for developing diffraction imaging [33]. The weak-beam imaging technique known in TEM refers specifically to the use of a low order diffracted beam with a large excitation error ($S_g$) for diffraction contrast imaging. This technique allows individual dislocations to be imaged as relatively intense, narrow peaks, positioned very close to the dislocation core [69, 70]. In electron diffraction, weak beams are formed by weak excitation or from weak reflections or obtained using a weak illumination as in the case of low-dose diffraction. An example of weak reflections is the (200) reflection of a III-V semiconductor with the zinc-blende structure and the use of (200) for DF imaging produces the so-called compositional contrast [71]. Low-dose electron diffraction is generally required for the study of molecular crystals. In electron diffraction imaging, a weak beam or multiple weak beams can be selected from the recorded DPs and by integrating the intensity of the selected beams for each pixel, a weak beam diffraction contrast image is formed. The advantage of this approach is two-fold. First, the electron dose per area (DPA) can be increased by several factors using a focused beam, while DPA in the TEM mode is limited by the requirement of a broad illumination. Consequently, the weak beam signal noise ratio can be greatly improved. Second, by recording DPs directly and processing them digitally, WBDI avoids the usual setup involved in diffraction contrast imaging and allows the use of virtual objective apertures of different shapes and sizes, as well as possibility to remove the background intensity, which is simply not possible with direct imaging in a TEM.

An early example of WBDI is the imaging of nanoscale charge ordering (CO) phase in $La_{1-x}Ca_xMnO_3$ (LCMO), where the only differences between the CO phase and the matrix are the modulations of atomic displacements [33]. The CO phase in LCMO with colossal magnetoresistance effect forms and disappears as the temperature is lowered [72]. The larger unit cell of the CO phase gives additional superlattice reflections (SLRs) in the diffraction pattern (Figure 5a). These SLRs can be used to image the CO phase, however, they are very weak compared to the fundamental reflections observed in both the CO phase and the matrix. In Figure 5, The intensity of the SLRs was measured from the recorded DPs and mapped according to the probe position. Figure 5b shows WBDIs at different temperatures as LCMO ($x = 0.45$) going through the phase transition. The color scale on the right corresponds to the ratio of the SLR peak intensity and the surrounding background. The areas with the intensities below the background level were flattened. Each map was obtained from the same single crystal domain in the LCMO sample ($x = 0.45$). This example demonstrates the combination of electron diffraction's ability to reveal the presence of ordered structures with scanning probe microscopy's ability to reveal those structures' real-space distribution [73].

A recent example of WBDI is the observation of defect nanodomains in defect-engineered metal–organic frameworks (MOFs) of UiO-66(Hf) exhibiting a blocky lamellar morphology [74]. The structure of MOFs, including the presence of defect nanodomains, is typically characterized by the bulk diffraction techniques, such as X-ray diffraction. However, knowledge about individual defects and hence defect interactions is lacking. Johnstone et al. [74] demonstrated that by forming multiple diffraction contrast images using low dose electron diffraction with DPA below the critical threshold of 10–20 $e^-$ $Å^{-2}$ for MOFs and the scanning technique, defect analysis in MOFs can be carried out at nm resolution (Figure 6). The observed nanodomains occur in a preferred orientation, extending perpendicular to (111) planes and sometimes perpendicular to other low index facets. The domain size can be both controlled by synthesis. The synthetic control of defect structures suggests a local molecular composition-controlled defect morphology. And the ability to characterize the microstructure in MOFs opens the possibility of extending the concept of defect engineering in hard materials to MOFs, enabled by electron diffraction imaging characterization.

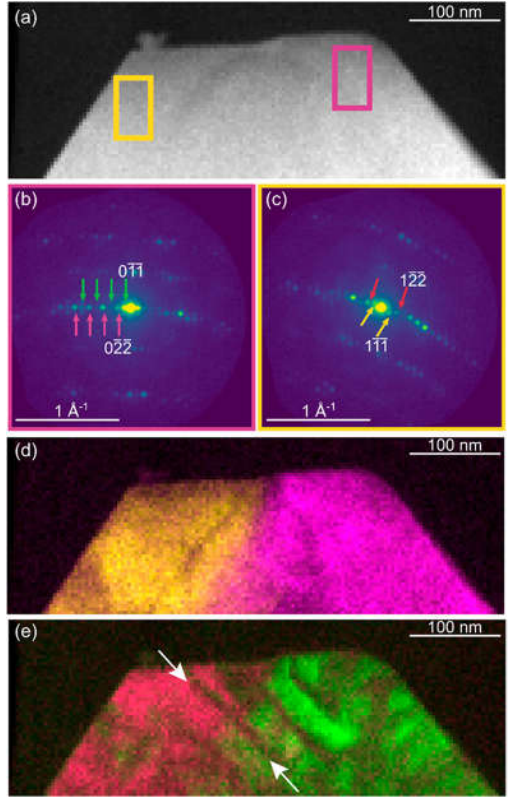

*Figure 6 Electron diffraction imaging analysis of a 6(Hf):5(BDC) UiO-66(Hf) particle with high defect density, containing a grain boundary. (a) ADF-STEM image indicating two marked regions, containing two different phases, where area averaged electron diffraction patterns (b, c) were obtained. The two phases differ by the presence of superlattice reflections. The two DPs also reveal a change in orientation between the left- and right-hand side of the particle. (d) Composite dark-field image formed using strong reflections. (e) Composite VDF image formed using weak superlattice reflections, green in b and red in c, revealing defect domains. (Figure taken from Ref. [74])*

### 3.2 Diffraction Pattern Clustering and Phase Identification

Clustering is broadly defined as the unsupervised classification of patterns [75]. It is a common statistical method, essential for data analysis [76]. The principle of DP clustering is same as image clustering, which is commonly employed in ML and computer vision. The aim of DP clustering is to identify similar DPs, average these similar patterns and obtain a correlation image where the distribution of DP clusters is mapped for the cluster-based imaging. The similarity of DPs can be quantified by the value of a predefined distance metric, for example, the normalized cross correlation coefficient (NCC)

$$\gamma = \frac{\sum_{x,y}\left\{\left[I_A(x,y)-\overline{I_A}\right]\cdot\left[I_B(x,y)-\overline{I_B}\right]\right\}}{\sqrt{\left\{\sum_{x,y}\left[I_A(x,y)-\overline{I_A}\right]^2\right\}\cdot\left\{\sum_{x,y}\left[I_B(x,y)-\overline{I_B}\right]^2\right\}}}, \qquad (1)$$

where $I_A(x,y)$ and $I_B(x,y)$ are intensities of the pixel $(x,y)$ in DP $A$ and $B$ respectively, and $\bar{I}_A$ and $\bar{I}_B$ are mean intensities of DP $A$ and $B$. The value of an NCC calculation between two DPs ranges from -1 to 1 with NCC = 1 indicating complete similarity. A cluster (which we will call as a group) is defined as one with all DPs belonging to the group having $\gamma$ values to be equal to or greater than a fixed, pre-defined, threshold value. This fixed threshold is called 'correlation threshold' (CT). The number of groups generally increases with the CT value with finer distinctions as CT increases. Thus, CT also defines the resolution of clustering analysis.

Multiple clustering techniques, such as the K-means clustering, agglomerative hierarchical clustering and density-based clustering, have been developed. K-means is an iterative cluster technique that targets to find a user-specified number of clusters. Rather than fixing the number of clusters, Zuo developed a modified K-means algorithm for a fast clustering of experimental DPs using random seeds [38, 77]. A user provided CT value is used to define the initial number of clusters and DPs belonging to the cluster. The DPs in each identified cluster are then averaged to obtain a cluster template, which is further used to find similar DPs. Figure 7a-b show an example of DP clustering analysis of an Au nano-disk using Zuo's algorithm. Among different DPs being identified, four are selected and shown in Figure 7c-f.

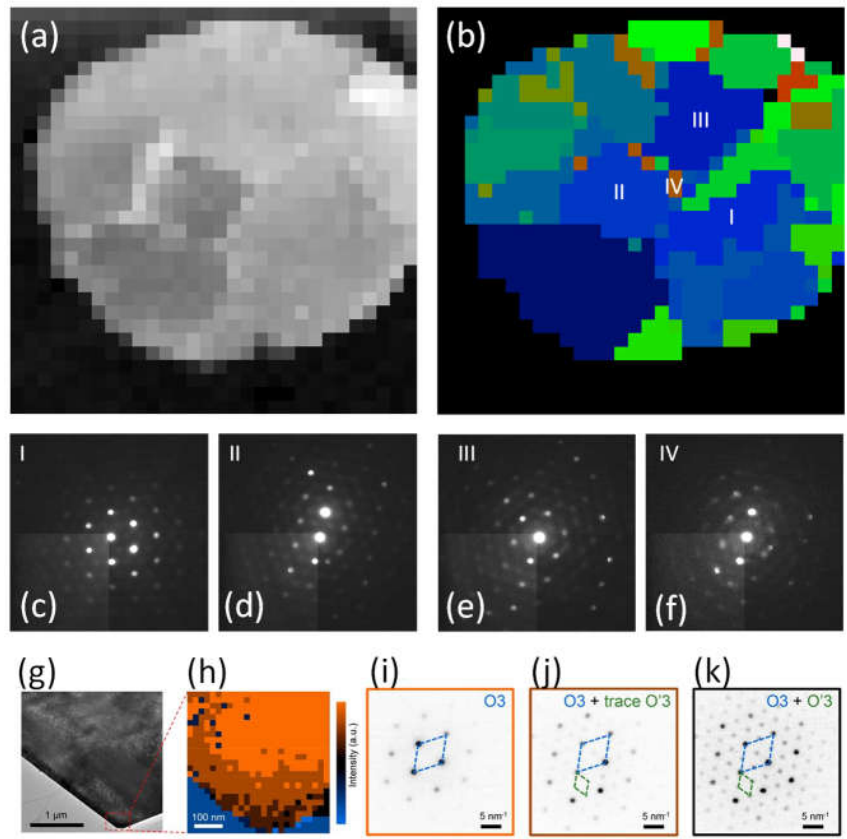

*Figure 7 DP clustering. a) a virtual ADF image reconstructed from a 30x30 diffraction scan over an Au nanodisk with a 7 nm step size for the image size of 210 nm, b) DPCI result with each color representing one DP cluster, a total of 37 clusters were identified, and c) to f) cluster averaged DPs for clusters marked*

*as I to IV in b). g) TEM image of a section of an electrodeposited platelet of $NaCoO_2$. h) A spatial phase map by SEND from the area (marked by red dotted border in g) shows that the central major part of the platelet is O3 whereas a thin layer of O′3 is formed at the edge of the platelets (~60 to 120 nm). Diffraction patterns are acquired from i) center (light brown square), j) middle (dark brown square), and k) edge of the platelet (black square). (Figures g to k are taken from Ref.[78])*

The K-means based clustering algorithms treat a DP as a whole pattern. Eggeman et al. reported a form of multivariate analysis, called non-negative matrix factorization (NMF) to decompose each diffraction pattern into the constituent components [79]. These components could be associated with a specific phase or orientation. A DP is a linear sum of component patterns, and each component is weighted by a loading coefficient. The number of component patterns is determined by trial-and-error, while the component patterns are kept positive. Recently, the NMF algorithm has been compared with the virtual dark-field (VDF) method for image segmentation, in which diffraction spots belong to the same crystal are identified by correlating their VDF images and the decomposed DPs are then used for image segmentation. It was demonstrated that both strategies can be used for nanocrystal segmentation without prior knowledge of the crystal structures present, but also that segmentation artefacts can arise and must be considered carefully [80].

A major application of clustering analysis is to identify phases and their distribution in a nanomaterial. Figure 7g-k show an example. To understand the physical distribution of the O3 and O′3 phases in $NaCoO_{2[78]}$, a platelet from an electrodeposited sample is analyzed with a 5-nm nanobeam via SEND over a 500 × 500-nm area across the edge of the platelet (shown in Figure 7g). The phase map of this area (Figure 7h) shows the central major part of the platelet is O3 and a small fraction of O′3 is formed at the edge of the platelet (~60 to 120 nm). DPs acquired from the center (light brown square, Figure 7i), middle (dark brown square, Figure 7j), and edge (black square, Figure 7k) of the platelet show a gradual transition of O3 (blue dotted) to O′3 (dark green dotted) near the edge, with the phase fraction of O′3 increasing markedly near the edge. The phase evidences from electron diffraction are consistent with the phase diagram interpretation and enable the controlled synthesis of layered sodium transition metal oxides [78].

### 3.3 Automatic crystal diffraction pattern indexing and orientation mapping

Indexing of a large DP dataset collected from a nanocrystalline material requires a fast and robust approach, which is only practicable through automation. Various automatic indexing schemes have been proposed and developed for X-ray and electron diffraction. Most of these schemes are based on comparison of experimental and calculated diffraction patterns [23, 81-84]. For electron diffraction, the successful approaches for automated crystal orientation mapping (ACOM) so far are based on the template matching [23] and correlations between diffraction peaks [85]. In template matching, the crystal orientation (and phase) is determined from the best fit between the experimental pattern and the pre-calculated diffraction pattern templates.

In preparing calculated diffraction patterns, two considerations are taken; one is the distance to the Ewald sphere as measured by the excitation error $S_g$. The reflections with $S_g$ smaller than a cutoff value are selected and their intersection with the detector plane determines the geometry of the diffraction pattern (Figure 8a). Second is the diffraction spot intensity. The diffraction intensity in the kinematical approximation for a perfect crystal of thickness $t$ is given by

$$I_g(t, S_g) = \pi^2 \lambda^2 \frac{\sin^2(\pi S_g t)}{(\pi S_g t)^2} |U_g|^2 t^2 . \qquad (2)$$

Where $U_g$ is the electron structure factor in the unit of Å$^{-2}$. In practice, the crystal thickness is often unknown, the measured peak intensity is integrated over a range of $S_g$. Multiple scattering also makes the intensity calculations based the kinematical approximation unreliable. Nonetheless, the inclusion of diffraction intensity in diffraction pattern indexing is often useful, especially in the near-zone axis situations. In such cases, only an estimate of the diffraction peak intensity can be made for irregular shaped crystals. Often an ad-hoc formula is used to assign the spot peak intensity. The formula below is an example

$$P(g) = A \log \left[ \frac{|U_g|^2}{1 + (S_{g,\min} / \lambda |U_g|)^2} \right] - B(g) \qquad (3)$$

where $U_{hkl}$ is the electron structure factor in the unit of Å$^{-2}$, $S_{g,\min}$ is smallest excitation error of reflection g for an incident convergent beam, A is a scaling constant and $B(g)$ is a length dependent correction function for the intensity of high order reflections. The log function is introduced to scale diffraction intensities. The oscillatory term involving $S_g t$ in equation is integrated out over a range of values.

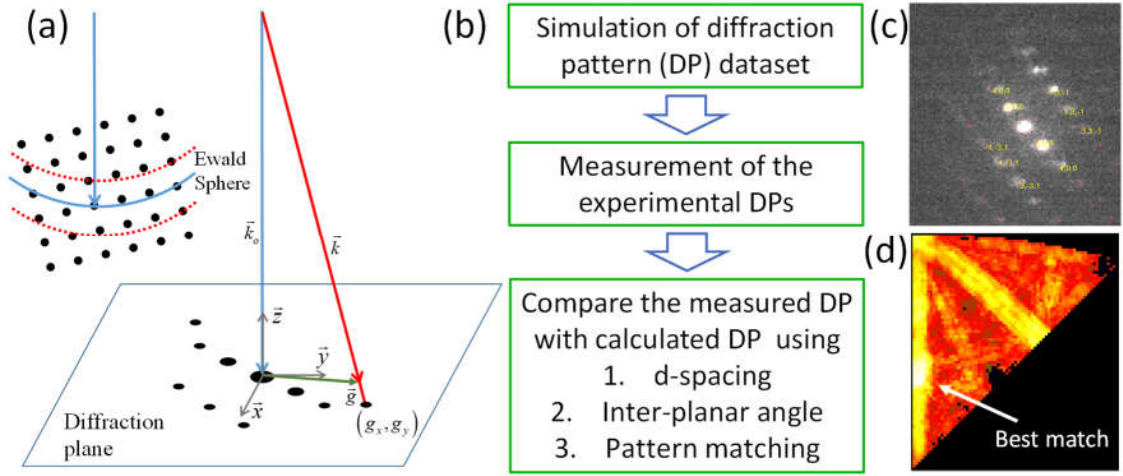

*Figure 8 Principles of automatic electron diffraction indexing. a) Transmission electron diffraction geometry and Ewald sphere construction (inset), b) Major steps involved in automatic diffraction pattern indexing, c) an example of indexed DP from TiN and d) index merit map showing the peak for best match.*

The matching between experimental and simulated DPs involves several steps. The details of finding the match vary among different implementations [85, 86]. Figure 8b summarizes the essential steps involved in indexing a DP as described by Meng and Zuo [85] together with an indexed example (Figure 8c). The best match is determined using a quality of matching index (QMI). Figure 8c shows the examples of an indexed diffraction pattern of TiN and the corresponding QMI map with the best match indicated by the arrow. The number of DPs to be indexed can be significantly reduced by combining with DP clustering method described in the previous section.

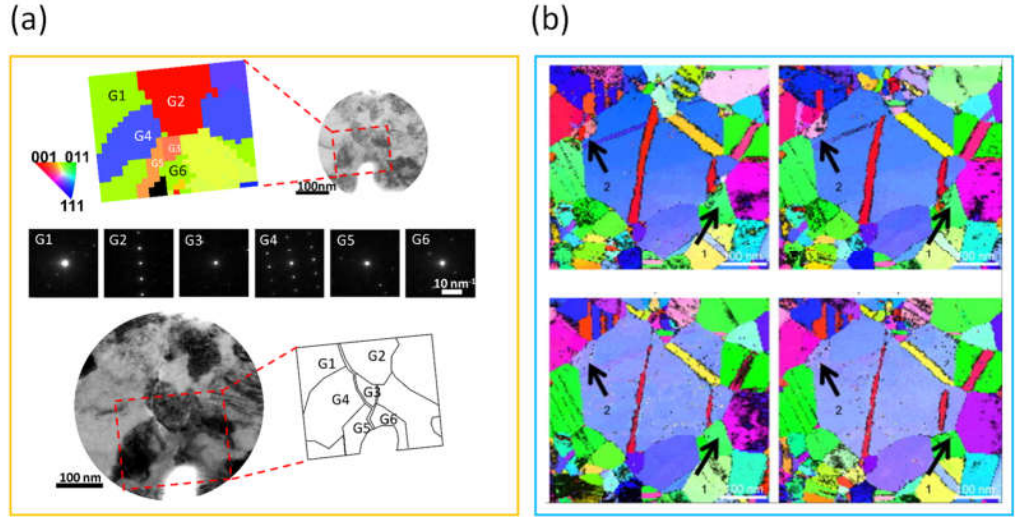

*Figure 9 Applications of OIM in nanocrystalline materials. a) Intergranular fracture mode of TiN hard coating, from top to bottom, orientation map of a FIB fabricated thin section with a notch, the map is obtained from the boxed region using SEND; the diffraction pattern for each grain; image taken after fracture showing the crack growth path. (Figure provided by Yang Hu, UIUC) b) Evolution of grains in sputter-coated Au film during tensile test a revealed ACOM-TEM (From Ref. [87]).*

A major application of automatic DP indexing is crystal orientation and phase mapping. Applications of TEM based orientation mapping, for example, have been demonstrated for the characterization of copper interconnect lines fabricated by the damascene process in microelectronic devices [88], nanocrystalline Cu [87], crack growth in nanocrystalline TiN thin-films [89] (Figure 9a) and in-situ deformation of nanocrystalline Au [90] (Figure 9b). A major advantage of TEM versus SEM based orientation mapping is the spatial resolution. Reliable EBSD analysis of nanocrystalline materials with grain sizes below 30–50 nm is difficult. Compared to TEM or STEM imaging-based nanostructure analysis, orientation mapping provides a quantitative identification of grains and their orientations and direct orientation information, while obtaining such information from electron images for many grains is very tedious.

### 3.4 Local Crystal Symmetry Determination and Mapping

CBED is a well-established and powerful technique for determining crystal point and space group symmetry at nanoscale [91-95]. Recent studies have demonstrated that local crystal symmetry and polarization nanodomains can be determined using scanning CBED [94, 96, 97]. By rastering the convergent electron probe over a region of the crystal, symmetry fluctuations and ferroelectric domains can be identified by changes in CBED patterns. A similar technique called STEM-CBED was used by Tsuda *et al.* for the same purpose [98, 99].

To quantify the symmetry of the CBED patterns, Kim and Zuo [94] used the NCC values (eqn. (1)) to compare a pair of diffraction discs of symmetry relations. In CBED patterns with multiple pair of symmetry related disks, the γ values of each pair of discs can be weighted to obtain the average with respect to intensity of the discs [100]. To deal with intensity redistribution due to a small crystal tilt or a slight tilt of the incident electron probe during scan, only part of the diffraction disk is used for symmetry quantification.

Figure 10 shows examples using the symmetry quantification algorithm for CBED patterns obtained from two sets of centrosymmetric and noncentrosymmetric crystals, Si, GaSb, SrTiO$_3$, and BaTiO$_3$ (regions outside the disks being quantified in the CBED pattern are being masked off). The observed symmetry and the measured γ values are marked on the CBED patterns.

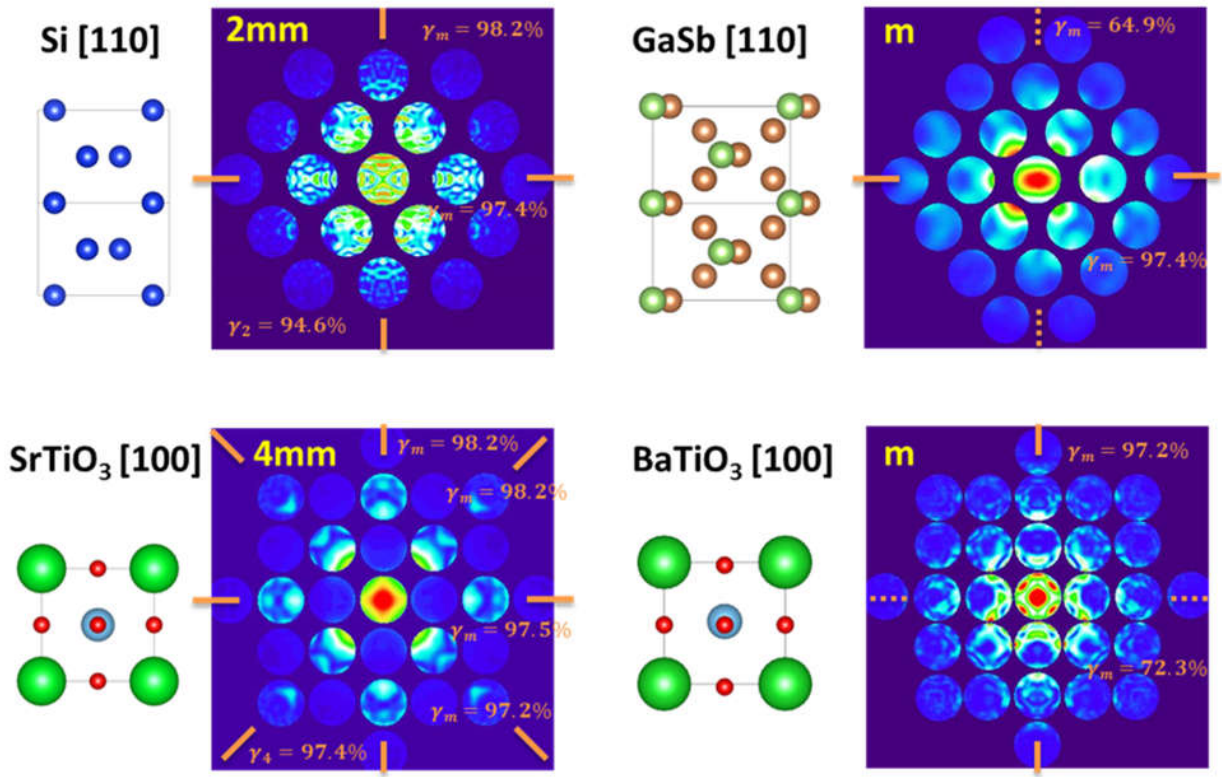

*Figure 10* Region averaged *CBED patterns obtained from centrosymmetric (left column) and noncentrosymmetric (right column) crystals* using SCBED.

The ability to determine local crystal symmetry makes SCBED very useful for the study of ferroelectric phase transitions. An example is the study of ferroelectric single crystal BaTiO$_3$. Phase transitions in BaTiO$_3$ are generally considered as a classic example of displacive soft-mode-type transitions, which are resulted from the instability of the lattice against a soft polar phonon at the center of the Brillouin zone [101-103]. Shao et al. revealed nanometer-sized regions of BaTiO$_3$ exhibiting mirror symmetry in accordance with the T-, O-, and R-symmetry in each of the corresponding phases at three different temperatures [97]. However, the regions exhibiting high symmetry are surrounded by regions with broken symmetries, with the latter in agreement with the findings by Tsuda *et al.* [104] and Tsuda & Tanaka [105]. The simultaneous presence of high and low symmetry regions can be explained by the coexistence of soft phonon modes and Ti ion off-center displacements, leading to both displacive and order-disorder character [106]. The other approach for the investigation of ferroelectric phase transitions is direct electron imaging at atomic resolution, which is particularly useful for observation of ferroelectric domain boundaries [96, 97, 100]. Compared to this approach, CBED provides high sensitivity to small atomic displacements associated with ferroelectric phase transition and the ability to examine complex crystals with lower symmetry phases, such as monoclinic polar domain boundaries [100].

### 3.5 Lattice strain analysis

The principle of determine lattice strain is to measure the two base reciprocal lattice vectors (g-vectors) of a 2D DP and use them to calculate strain with the help of two reference g-vectors [107]. The DP does not have to be at the exact zone axis orientation, if multiple diffracted beams are visible. In electron nanodiffraction using a parallel beam, lattice spacing can be directly measured from the diffraction peak positions [107-111]. A near-parallel illumination is set up to produce sharp, and well-defined, diffraction spots. The probe size ranges from several to tens of nm in diameter. For higher spatial resolution, a focused probe of small convergence angle is used with the probe size of ~1 nm. Several methods have been developed to detect the position of small diffraction disks in nanodiffraction patterns. Rouviere et al. [112] applied precession to the electron beam to make disk intensity uniform to improve the measurement precision. Other efforts made to locate non-uniform diffraction disks include the radial gradient maximization method [111], template matching using cross-correlation and its variations [109, 113]. Yuan et al. described a method for diffraction disk detection based on circular Hough transform (CHT) [25]. This method determines the position of diffraction disks based on their shapes, which are approximately similar. A ML based approach using convolutional neural network (CNN) can also be designed to measure diffraction disk position [114]. The experimental DPs can be divided into small sub-images; each contains one diffraction disk, the CNN is trained with simulated DPs or experimental DPs with known positions, and the trained CNN is then applied to recognize disk position. The advantage of using ML is that the dynamical effect can be included using simulated pattern, which solves the problem of identifying the appropriate template for disk detection.

The projected 2D strain $\varepsilon$ can be obtained using the 2D deformation matrix ($D$) obtained from two measure reciprocal vectors, $\vec{g}_1$ and $\vec{g}_2$. They can be taken as the basis vectors for the zone axis diffraction pattern or any two non-parallel vectors recorded in the diffraction pattern. Each vector is defined by its components along the x and y directions perpendicular to the zone axis. They give the following G-matrix:

$$G = \begin{pmatrix} g_{1x} & g_{2x} \\ g_{1y} & g_{2y} \end{pmatrix} \quad (4)$$

and $D$ is simply given by

$$D = \left(G^T\right)^{-1} G_o^T - I \quad (5)$$

where $T$ represents transverse, and $I$ is a unit diagonal matrix and $G_o^T$ is the transverse of the G matrix of the reference crystal. The strain and crystal rotation is obtained from $D$ using

$$\varepsilon = \frac{1}{2}(D+D^T) \text{ and } \omega = \frac{1}{2}(D-D^T). \quad (6)$$

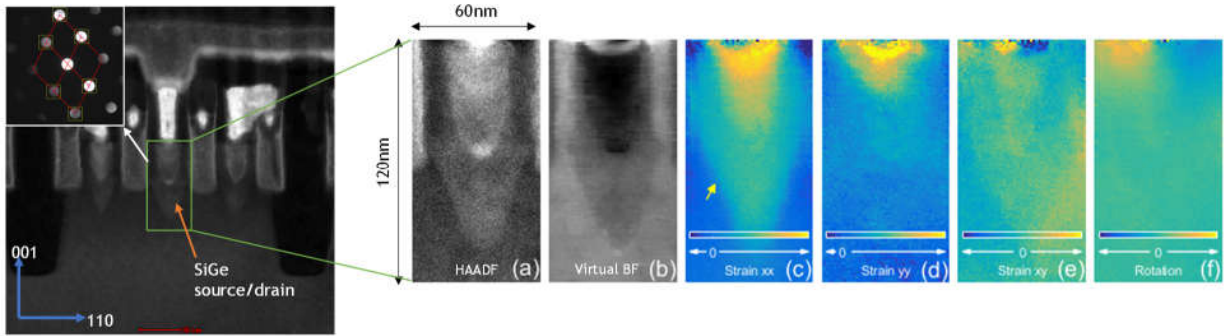

*Figure 11 Strain mapping of the SiGe source/drain region in a FinFET device. (a) ADF image of the scanned region. (b) Reconstructed virtual BF image of the scanned region shows that sample drift is well compensated. Maps of (c) $\varepsilon_{xx}$, (d) $\varepsilon_{yy}$, (e) $\varepsilon_{xy}$, and (f) rotation calculated from the scanning diffraction dataset at 1 nm spatial resolution. Left, an ADF image of the FinFET device and an example of the recorded DPs. The x direction of strain is defined as the [110] direction, and the y direction is [001].*

Lattice strain broadly affects the properties of materials. Large elastic strains can be generated by epitaxy or by localization under external loading. Strain can be spatially homogeneous or inhomogeneous. Nanoscale elastic strain engineering provides new possibilities for tuning the physical and chemical properties of a material, including electronic, mechanical, and electrochemical properties. Being able to determine lattice strain at nm resolution with high precision for a broad range of materials is thus a unique opportunity for electron diffraction imaging.

Figure 11 and Figure 12 highlight two examples. One is strain mapping of 3D FinFET device and the other is a determination strain fields around dislocations. In a FinFET device, by breaking the Si cubic symmetry and modifying its band structure, large device performance gain can be obtained by enhancing the carrier mobilities in the device channel. Starting from Intel's 22 nm process, the transistor has evolved to 3D tri-gate architecture with strain engineering from various approaches including strained epitaxial layers and stress films. Critical to the success of the further development of transistor technologies is the ability to analytically measure and quantify the strain. The strain maps in Figure 11 are obtained using the technique described by Yuan et al. [25]. Strain maps show clearly the interface between SiGe and Si substrate and the epitaxial nature of strain. The maximum of strain is reached near the top of source/drain where it is attached to metal contact. Traditional TEM strain techniques [110] that work for planar transistors, such as CBED and dark-field electron holography, do not apply to tri-gate transistors due to the complicated 3D tri-gate geometry. Electron nanodiffraction separates the device channel and the overlapped materials in DPs and enables strain measurement in the crystalline channel. The demonstrated our resolution and strain measurement precision in the FinFET device are 1 nm and $\sim 3 \times 10^{-4}$, respectively.

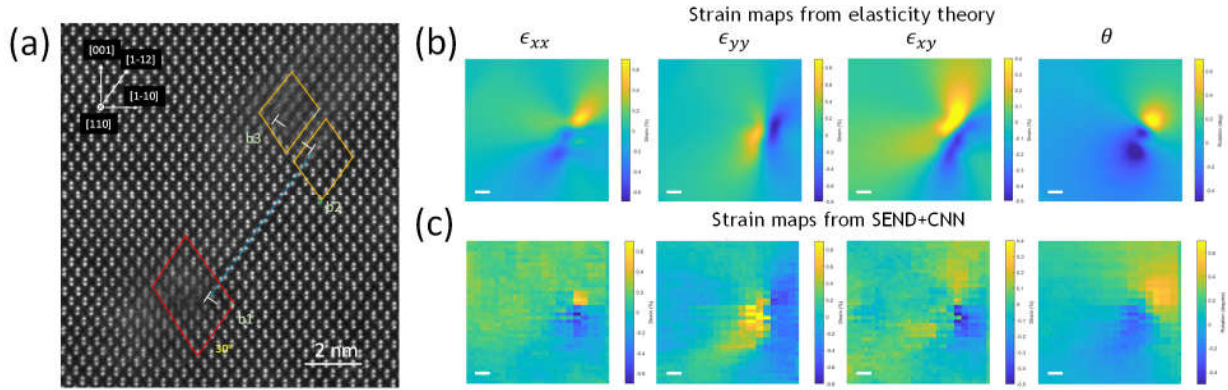

*Figure 12 Strain mapping of a complex defect in SiGe. (a) A HADDF image of the defect. (b) and (c) Strain maps $\epsilon_{xx}$, $\epsilon_{yy}$, $\epsilon_{xy}$, and rotation calculated by elasticity theory and measured experimentally using SEND. Here, x is along [1-10] and y along [001], and the scale bar is 4 nm. (From Yuan[115])*

Figure 12 shows the strain field of a misfit dislocation at the interface of two SiGe layers of different composition. The image reveals a stacking fault bounded by two 30° partial dislocations with Burgers vectors of $b_1 = 1/6[\bar{2}1\bar{1}]$ and $b_2 = 1/6[211]$. A perfect dislocation $b_3$ with a Burgers vector of $1/2[\bar{1}0\bar{1}]$ is found near $b_2$. The $b_2$ and $b_3$ are separated by 3 atomic layers. Figure 12b and c show the modelled and the measured strain maps, respectively. The measurements were based on the ML approached introduced by Yuan et al. [114]. This technique can achieve both high accuracy and high precision. The measured strain fields feature dipole-like features in $\epsilon_{xx}$ and rotation components, while complex features in $\epsilon_{yy}$ and $\epsilon_{xy}$ are observed. These features are in agreement with the modelled strain fields from three edge dislocations ($b_1$, $b_2$, and $b_3$ in Figure 12a) with a Gaussian blur to mimic the finite spatial resolution of SEND. The close match between the experimental and model strain maps demonstrates the good precision, sensitivity, and resolution of our strain mapping technique near the dislocation cores. While atomic resolution STEM imaging can resolve structures close to the dislocation core, the relatively weak strain fields extended away from the core requires high strain sensitivity of the diffraction approach.

### 3.7 Cepstral STEM using electron diffuse scattering

So far, we have focused on diffraction signals from Bragg diffraction, which is only part of the recorded DPs. Other signals that can be used for diffraction imaging come from diffuse scattering observed between Bragg spots [11, 116]. In crystals, a part of diffuse scattering is due to elastic scattering from structural imperfections, such as point defects, dislocations and stacking faults. Another source of diffuse scattering is partial ordering, for example short-range ordering in an alloy. The nature of diffuse scattering depends sensitively on its origin [29]. Thus, diffuse scattering signals can be used for determining crystal imperfections. However, the interpretation of diffuse scattering is a challenging problem in both X-ray and electron diffraction [117]. Another issue is that electron diffuse scattering is weak, their detection requires a detector with a large dynamic range and good sensitivity to low electron signals [30, 118]. The advance of direct electron detection technology now opens-up the possibility to map electron diffuse scattering intensities at nm spatial resolution.

In a nanodiffraction pattern obtained using a coherent electron beam, the diffuse scattering is more like laser speckles, in a way similar to fluctuations recorded in amorphous materials [68, 119]. Recently, Shao et al. introduced the difference Cepstrum method for the analysis of speckle patterns [120]. The difference Cepstrum ($dC_p$) is calculated according to

$$dC_p = \left|FT\left\{log\left[\frac{I(\vec{k})}{I_{avg}(\vec{k})}\right]\right\}\right| = \left|FT\{log[I(\vec{k})]\} - FT\{log[I_{avg}(\vec{k})]\}\right| \quad (7)$$

where $I_{avg}(\vec{k})$ represents intensity in the area averaged pattern, while $I(\vec{k})$ is the intensity in a single pattern from DPs collected over a ROI. Fourier transform of logarithmic of frequency signals is known as cepstrum analysis in signal processing [121, 122]. Cepstral analysis finds periodic structures in frequency spectra. The interpretation of $dC_p$ for electron diffraction is made based on the separation of the fluctuating part of the scattering potential ($U_1$) from the average scattering potential ($U_{avg}$), where $U_1$ varies with the electron probe position. The $U_{avg}$ represents an average over the region scanned by the electron probe, which for a randomly disordered crystal describes the periodic scattering potential. Diffraction by $U_{avg}$ gives $I_{avg}(\vec{k})$. Shao *et al.* show that $dC_p$ approximately corresponds to the autocorrelation function, or Patterson function (PF) [123], of the distortive part of scattering potential ($U_1$) in a thin sample.

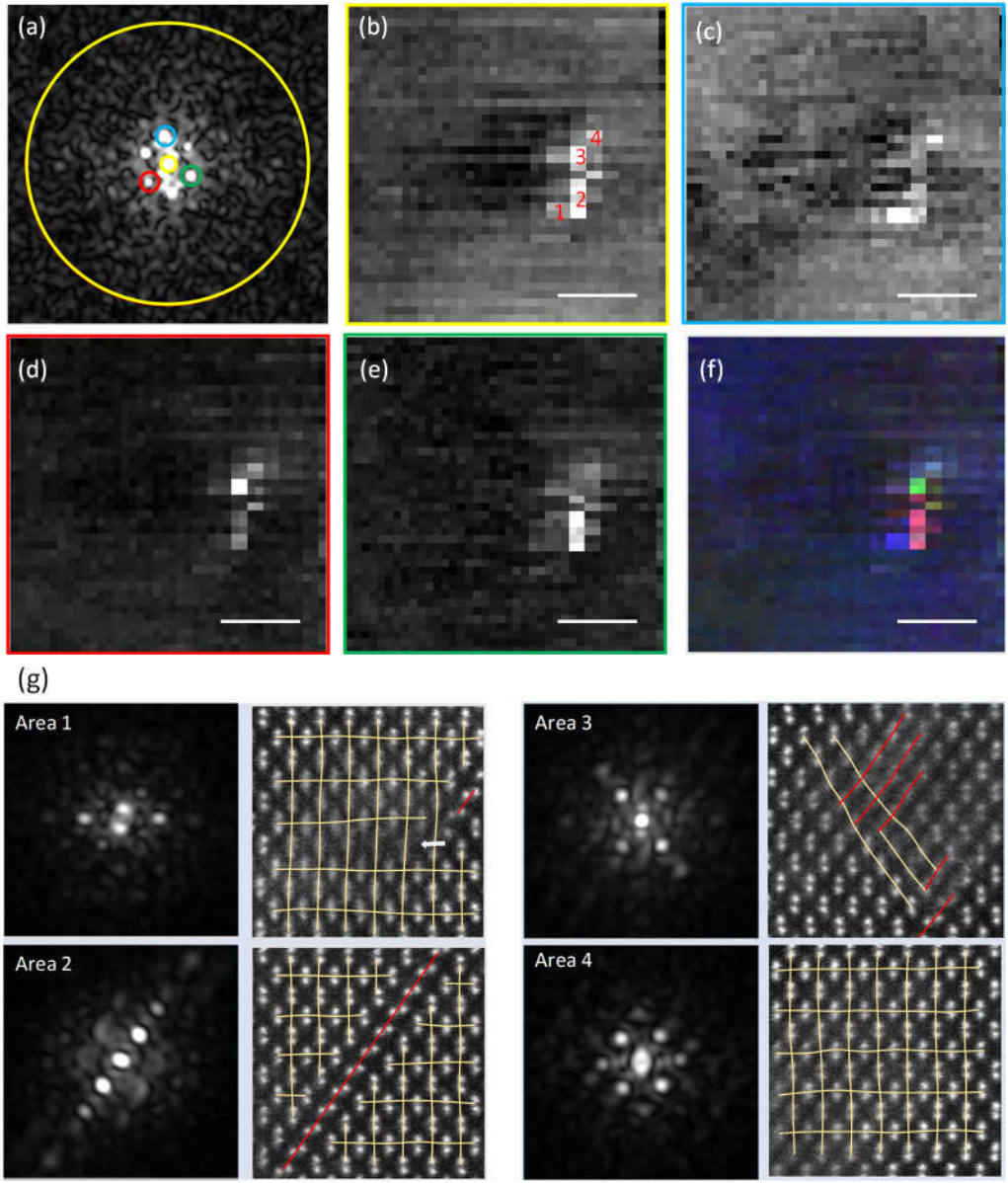

*Figure 13 Cepstral STEM imaging of a dislocation core. (a) Cepstral data where intensity within the marked yellow circles is integrated to form Cepstral ADF image of (b). (c) DF image formed with intensities within the blue circle, (d) and (e) within the red and green circles, respectively. (f) The RGB image formed using (c), (d) and (e). The scale bar is 10 nm. (g) Differential Cepstral patterns from 4 areas of dislocation core in (b) and comparison with corresponding atomic resolution HAADF-STEM images.*

The sensitivity of $dC_p$ to the distortive part of potential is demonstrated in Figure 13, in which the dislocation core imaged in Figure 12a is analyzed. Two cepstral STEM imaging modes are demonstrated: the first is ADF imaging (Figure 13b) and the second is DF imaging using the quefrency signals (Figure 13c-e). Here quefrency is the variable in cepstrum with the unit of length. The ADF cepstral STEM image intensity is related to the amount of diffuse scattering recorded in DPs. In this case strong diffuse scattering is detected at the dislocation core (Figure 13b). The DF cepstral STEM images show the local where a particular quefrency signal is detected. Based on the DF images, four areas (1-4) are identified in the dislocation core. Each of these four areas gives distinct $dC_p$ pattern, which can be correlated directly with lattice distortions detected at the dislocation by atomic resolution HAADF image (Figure 13g). The quefrency signals are related to the Patterson function of diffuse scattering, arising from the distorted part of scattering potential [120].

The example here demonstrates ways that quefrency signals detected by Cepstral analysis can be used to form Cepstral STEM images of lattice change in the highly distorted regions. The information obtained by Cepstral STEM image complements the Bragg reflection-based analysis of strain by providing information about severely distorted regions of crystals.

## 4. Future Opportunities

Since the first TEM was demonstrated by Max Knoll and Ernst Ruska in 1931, the field of electron microcopy has rapidly grown with innovations in the microscope technology, electron optics and methodology. However, quantitative electron imaging is a rather recent development, starting with HREM [124, 125] and continuing with aberration corrected STEM [126]. Electron diffraction provides highly quantifiable signals. Thus, we expect the development of electron diffraction imaging will accelerate the trend toward quantitative electron microscopy. Figure 14 illustrates the roles and potentials of electron diffraction imaging in the large scheme of materials characterization, where the imaging modes reviewed here provide information from atomic structure to microstructure. It should be emphasized that the type of materials that can be characterized using diffraction imaging approach include both hard and soft materials [127].

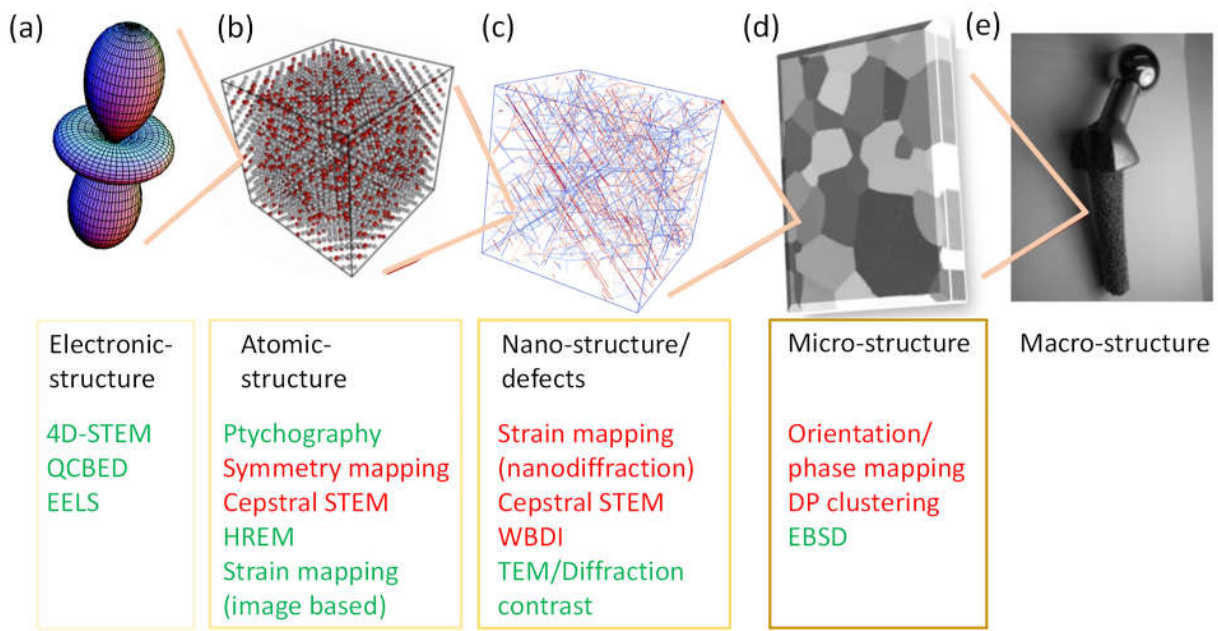

*Figure 14 Schematic illustration of multi-length scale materials structure characterization and various electron microscopy techniques that have been developed to meet the characterization needs. Techniques*

*in red are reviewed here based on electron diffraction imaging. (image credits, (b), (d) and (e) are from Ref. [128], (c) is from Ref.[129] )*

The large amount of data is the essential aspect of data-driven electron microscopy and a major advantage over conventional electron microscopy. However, the amount of data collected can be ~100k times larger or more than a typical electron image (for a detector of 512×512 pixels, for example). Currently, DPs are collected for every probe position in a rectilinear scan over a ROI and the field-of-view (FoV) of scan is rather limited for a fixed data size. The dense sampling in such scan creates many redundancies in the recorded DPs, as homogeneous regions produce similar DPs. Thus, one opportunity in electron diffraction imaging is to explore the minimum number of sampling points. Along this line, the concept of compressive sensing (CS) [130], which has been explored in direct STEM imaging, is a possible solution. The key is to significantly reduce the number of sampling points, while maintaining all the important features in an image [131-135]. A benefit of such approach will be the reduced data size, which can be used to expand the FoV. Another benefit is the reduction in the overall electron dose, which will help with the characterization of beam sensitive materials.

Automatic DP indexing in nanocrystalline materials continues to be a challenge, especially in indexing DPs far away from major zone axes with few diffraction spots and DPs collected from overlapping grains. Both can lead to ill-defined maximum or multi-peaks in the QMI map. Recent progress in ML provides a possible way forward. The progress made in DP clustering and segmentation provides a basis to integrate automatic DP indexing for a possible solution to this challenging problem. Another option to be explored is the use of ML.

We have introduced the principles for using electron diffuse scattering for STEM imaging. We show that the Cepstral difference can be related to the Patterson function of diffuse scattering, arising from the distorted part of scattering potential. Using examples of dislocations in SiGe and high entropy alloy, Shao et al. [120] demonstrated ways that harmonic signals detected by Cepstral analysis can be used to form Cepstral STEM images. The Cepstral STEM image contrast is compared with that of regular STEM and Bragg reflection-based strain mapping. The results show that these image modes are complementary, with Cepstral STEM providing the critical information about severely distorted regions of crystals. The interpretation of volume averaged diffuse scattering for the study of disordered crystals is generally difficult and often requires the help of modeling. Being able to directly image fluctuations in diffuse scattering at nm resolution and detect harmonic signals thus provides a new way for studying crystal disorder.

**Acknowledgments**
This writing of this review was supported by Ivan Racheff Professorship of Materials Science and Engineering, University of Illinois and a Strategic Research Initiative grant from Grainger College of Engineering, University of Illinois.